\documentstyle[prd,aps,floats]{revtex}

\def\be{\begin{equation}}
\def\ee{\end{equation}}
\def\bea{\begin{eqnarray}}
\def\eea{\end{eqnarray}}

\begin{document}
\preprint{SUSX-TH-97-022, SUSSEX-AST 97/11-1, PU-RCG/97-20, gr-qc/9711068}
\draft

%
%
\input epsf
\renewcommand{\topfraction}{0.99}
\renewcommand{\bottomfraction}{0.99}
\twocolumn[\hsize\textwidth\columnwidth\hsize\csname 
@twocolumnfalse\endcsname

\title{Exponential potentials and cosmological scaling solutions} 
\author{Edmund J.~Copeland}
\address{Centre for Theoretical Physics, University of Sussex, Falmer, 
Brighton BN1 9QH,~~~U.~K.}  
\author{Andrew R.~Liddle}
\address{Astronomy Centre, University of Sussex, Falmer, Brighton BN1
9QH,~~~U.~K.}  
\author{David Wands}
\address{School of Computer Science and Mathematics, University of 
Portsmouth, Portsmouth PO1 2EG,~~~U.~K.}  
\date{\today} 
\maketitle
\begin{abstract}
We present a phase-plane analysis of cosmologies containing a barotropic 
fluid with equation of state $p_\gamma = (\gamma-1) \rho_\gamma$, plus a 
scalar field $\phi$ with an exponential potential $V \propto \exp(-\lambda 
\kappa \phi)$ where $\kappa^2 = 8\pi G$. In addition to the well-known 
inflationary solutions for $\lambda^2 < 2$, there exist scaling solutions 
when $\lambda^2 > 3\gamma$ in which the scalar field energy density tracks 
that of the barotropic fluid (which for example might be radiation or dust). 
We show that the scaling solutions are the unique late-time attractors 
whenever they exist. The fluid-dominated solutions, where 
$V(\phi)/\rho_\gamma \rightarrow 0$ at late times, are always unstable 
(except for the cosmological constant case $\gamma = 0$).
The relative energy density of the fluid and scalar field depends on the 
steepness of the exponential potential, which is constrained by 
nucleosynthesis to $\lambda^2 > 20$. We show that standard inflation models 
are unable to solve this `relic density' problem.
\end{abstract}

\pacs{PACS numbers: 98.80.Cq\\ 
Preprint SUSX-TH-97-022, SUSSEX-AST 97/11-1, PU-RCG/97-20, gr-qc/9711068}

\vskip2pc]

\section{Introduction}

Scalar fields have come to play a central role in current models of
the early universe. The self-interaction potential energy density of
such a field is undiluted by the expansion of the universe and hence
can act like an effective cosmological constant driving a period of
inflation. The detailed evolution is dependent upon the specific form
of the potential $V$ as a function of the scalar field's expectation
value $\phi$.

A common functional form for the self-interaction
potential is an exponential dependence upon the scalar field. It is 
to be found in higher-order \cite{whitt84} or higher-dimensional
gravity theories \cite{KK}. In string or Kaluza--Klein type models the moduli
fields associated with the geometry of the extra dimensions may have
effective exponential potentials due to curvature of the internal
spaces, or the interaction of moduli with form fields on the internal
spaces. Exponential potentials can also arise due to non-perturbative
effects such as gaugino condensation~\cite{gaugino}. 

The possible cosmological roles of exponential potentials have been
investigated before, but almost always as a means of driving a period
of cosmological inflation~\cite{LM85,pplane}. This requires potentials that
are much flatter than those usually found in particle physics
models. The purpose of this paper is to emphasize that scalar
fields with exponential potentials may still have important
cosmological consequences even if they are too steep to drive a period
of inflation~\cite{Wetterich,texas,FJ}. We will present a phase-plane
analysis to show that scalar fields with exponential potentials
contribute a non-negligible energy density at nucleosynthesis unless
they are unusually steep. This `relic density' problem is not
alleviated by standard models of inflation.

\section{Autonomous phase-plane}

We will consider a scalar field with an exponential potential energy
density $V=V_0\exp(-\lambda\kappa\phi)$ evolving in a spatially-flat
Friedmann--\-Robertson--\-Walker (FRW) universe containing a fluid with 
barotropic equation of state
$p_\gamma=(\gamma-1)\rho_\gamma$, where $\gamma$ is a constant,
$0\leq\gamma\leq2$, such as radiation ($\gamma=4/3$) or dust
($\gamma=1$). The evolution equations for a spatially-flat FRW 
model with Hubble parameter $H$
are 
\bea
\dot{H} & = & - \, {\kappa^2\over2} \left( \rho_\gamma + p_\gamma +
\dot\phi^2 \right) \ , \\
\dot\rho_\gamma &=& -3H(\rho_\gamma+p_\gamma) \ , \\
\ddot\phi &=& - 3H\dot\phi - {dV\over d\phi} \ ,
\label{eq3}
\eea
subject to the Friedmann constraint
\be
\label{Friedmann}
H^2 = {\kappa^2\over3} \left( \rho_\gamma + {1\over2}\dot\phi^2 +
V \right) \ ,
\ee
where $\kappa^2\equiv8\pi G$. The total energy density of a
homogeneous scalar field is $\rho_\phi=\dot\phi^2/2+ V(\phi)$.

\begin{table*}[t]
\begin{center}
\begin{tabular}{|c|c|c|c|c|c|}
$x$ & $y$ & Existence & Stability & $\Omega_\phi$ 
 & $\gamma_\phi$ \\ 
\hline 
\hline 
0 & 0 & All $\lambda$ and $\gamma$ & Saddle point 
for $0 < \gamma < 
2$ &   0 & Undefined \\
\hline
 1 & 0 & All $\lambda$ and $\gamma$ & Unstable node for $\lambda <
 \sqrt{6}$ & 1 & 2 \\ 
 & & & Saddle point for $\lambda > \sqrt{6}$ & & \\
\hline
-1 & 0 & All $\lambda$ and $\gamma$ & Unstable node for $\lambda >
-\sqrt{6}$ & 1 & 2 \\ 
 & & & Saddle point for $\lambda < -\sqrt{6}$ & & \\
\hline
$\lambda/\sqrt{6}$ & $[1-\lambda^2/6]^{1/2}$ & $\lambda^2 < 6$ &
Stable node for $\lambda^2 < 3\gamma$ & 1 & $\lambda^2/3$ \\ 
 & & & Saddle point for $3\gamma < \lambda^2 < 6$ & & \\
\hline
$(3/2)^{1/2} \, \gamma/\lambda$ & $[3(2-\gamma)\gamma/2\lambda^2]^{1/2}$ 
& $\lambda^2 > 3\gamma$ & Stable node for $3\gamma < \lambda^2 < 
24 \gamma^2/(9\gamma -2)$ & $3\gamma/\lambda^2$ & $\gamma$ \\ 
 & & & Stable spiral for $\lambda^2 > 24 \gamma^2/(9\gamma -2)$ & & \\
\end{tabular}
\end{center}
\caption[crit]{\label{crit} The properties of the critical points.}
\end{table*}

We define
\be
x \equiv {\kappa\dot\phi \over \sqrt{6}\,H} \quad ; \quad
y \equiv {\kappa\sqrt{V} \over \sqrt{3}\,H} \ .
\ee
The evolution equations can then be written as
a plane-autonomous system:
\bea
\label{eomx}
x' & = & -3x + \lambda \sqrt{{3\over2}} y^2
 + {3\over2} x \left[ 2x^2 + \gamma \left( 1 - x^2 - y^2 \right)
\right] \ , \\
\label{eomy}
y' & = & - \lambda \sqrt{{3\over2}} xy
 + {3\over2} y \left[ 2x^2 + \gamma \left( 1 - x^2 - y^2 \right)
\right] \ ,
\eea
where a prime denotes a derivative with respect to the logarithm of
the scale factor, $N\equiv\ln(a)$,
and the constraint equation becomes
\be
{\kappa^2\rho_\gamma \over 3H^2} + x^2 + y^2 = 1 \ .
\ee

Note that from the constraint equation we have
\be
\Omega_\phi \equiv {\kappa^2 \rho_\phi \over 3H^2} = x^2 + y^2 \ .
\ee
This is bounded, $0\leq x^2+y^2\leq1$, for a non-negative fluid
density, $\rho_\gamma\geq0$, and so the evolution of this system is
completely described by trajectories within the unit disc.  The lower
half-disc, $y<0$, corresponds to contracting universes. As the system
is symmetric under the reflection $(x,y)\to(x,-y)$ and time reversal
$t\to-t$, we only consider the upper half-disc, $y\geq0$ in the
following discussion.

The effective equation of state for the scalar field at any point
is given by
\be
\gamma_\phi \equiv {\rho_\phi+p_\phi \over \rho_\phi} 
 = {\dot\phi^2 \over V + \dot\phi^2/2}
 = {2x^2 \over x^2 + y^2}
\ee
Fixed points at finite values of $x$ and $y$ in the phase-plane
correspond to solutions where the scalar field has a barotropic
equation of state and the scale factor of the universe evolves as
$a\propto t^p$ where $p={2/3\gamma_\phi}$.

Depending on the values of $\gamma$ and $\lambda$, we have up to
five fixed points (critical points) where $x'=0$ and
$y'=0$ which are listed in Table~I. A full analysis of the stability is given 
in the Appendix. 

Two of the fixed points ($x=\pm1$,
$y=0$) correspond to solutions where the constraint
Eq.~(\ref{Friedmann}) is dominated by the kinetic energy of the
scalar field with a stiff equation of state, $\gamma_\phi=2$. As
expected these solutions are unstable and are only expected to be
relevant at early times.

More surprisingly, however, we find that the barotropic
fluid dominated solution ($x=0$, $y=0$) where $\Omega_\phi=0$ is {\em
unstable} for all values of $\gamma>0$. We will discuss the critical
case where $\gamma=0$ later. But for any $\gamma>0$, and however steep
the potential (i.e.~whatever the value of $\lambda$), the energy density of
the scalar field {\em never} vanishes with respect to the other matter
in the universe.

We are left with only two possible late-time attractor solutions. One
of these is the well-known scalar field dominated solution
($\Omega_\phi=1$) which exists for sufficiently flat potentials,
$\lambda^2<6$. The scalar field has an effective barotropic index
$\gamma_\phi=\lambda^2/3$ giving rise to a power-law inflationary
expansion~\cite{LM85} ($\ddot{a}>0$) for
$\lambda^2<2$. Previous phase-plane
analyses~\cite{pplane} have shown that a wide class of
homogeneous vacuum models approach the spatially-flat FRW model for
$\lambda^2<2$. We have shown that this scalar field dominated solution
is a late-time attractor in the presence of a barotropic fluid when we
have $\lambda^2<3\gamma$.

However for $\lambda^2>3\gamma$ we find a different late-time attractor
where neither the scalar-field nor the barotropic fluid entirely
dominates the evolution. Instead we have a scaling solution where the
energy density of the scalar field remains proportional to that of the
barotropic fluid with $\Omega_\phi=3\gamma/\lambda^2$. This solution was
first found by Wetterich~\cite{Wetterich} and shown to be the global
attractor solution for $\lambda^2>3\gamma$ in Ref.~\cite{texas}.

\begin{figure}[t]
\centering 
\leavevmode\epsfysize=5cm \epsfbox{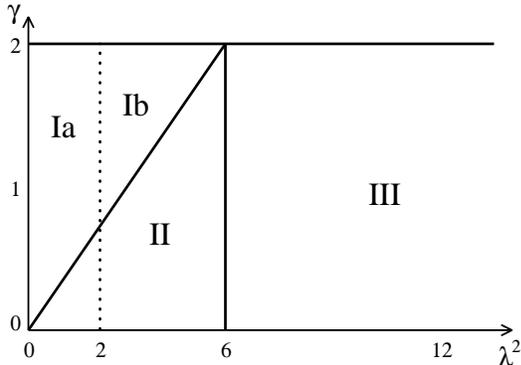}\\ 
\caption[regions]{\label{regions} Regions of $(\gamma,\lambda)$ parameter 
space, as identified in the text. Solutions to the left of the dotted line 
are inflationary.}
\end{figure}

\begin{figure}[ht!]
\centering 
\leavevmode\epsfysize=5cm \epsfbox{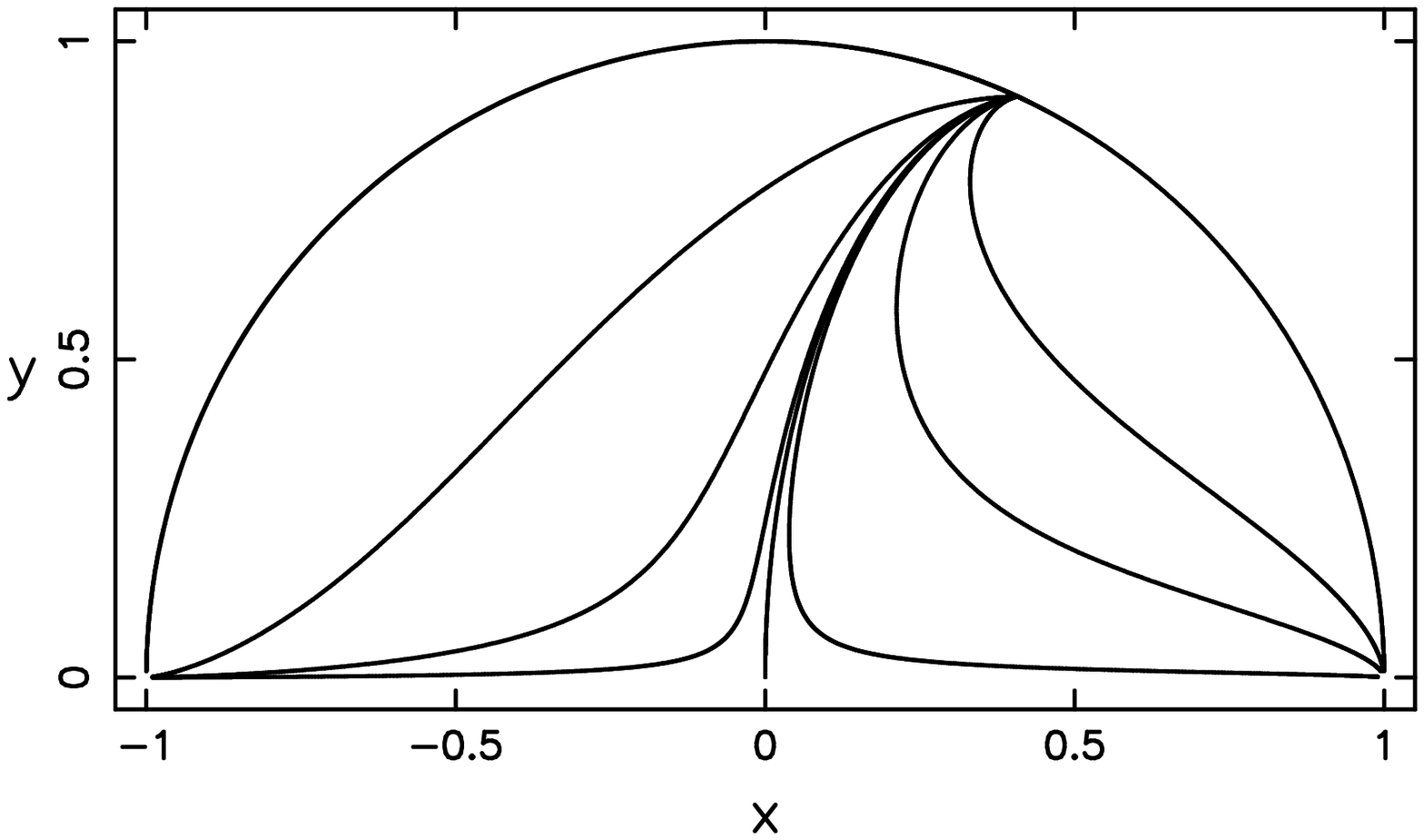}\\ 
\caption[phase1]{\label{phase1} 
The phase plane for $\gamma = 1$, $\lambda = 1$. The late-time attractor
is the scalar field dominated solution with $x=\sqrt{1/6}$, $y=\sqrt{5/6}$.}
\vspace*{12pt}
\leavevmode\epsfysize=5cm \epsfbox{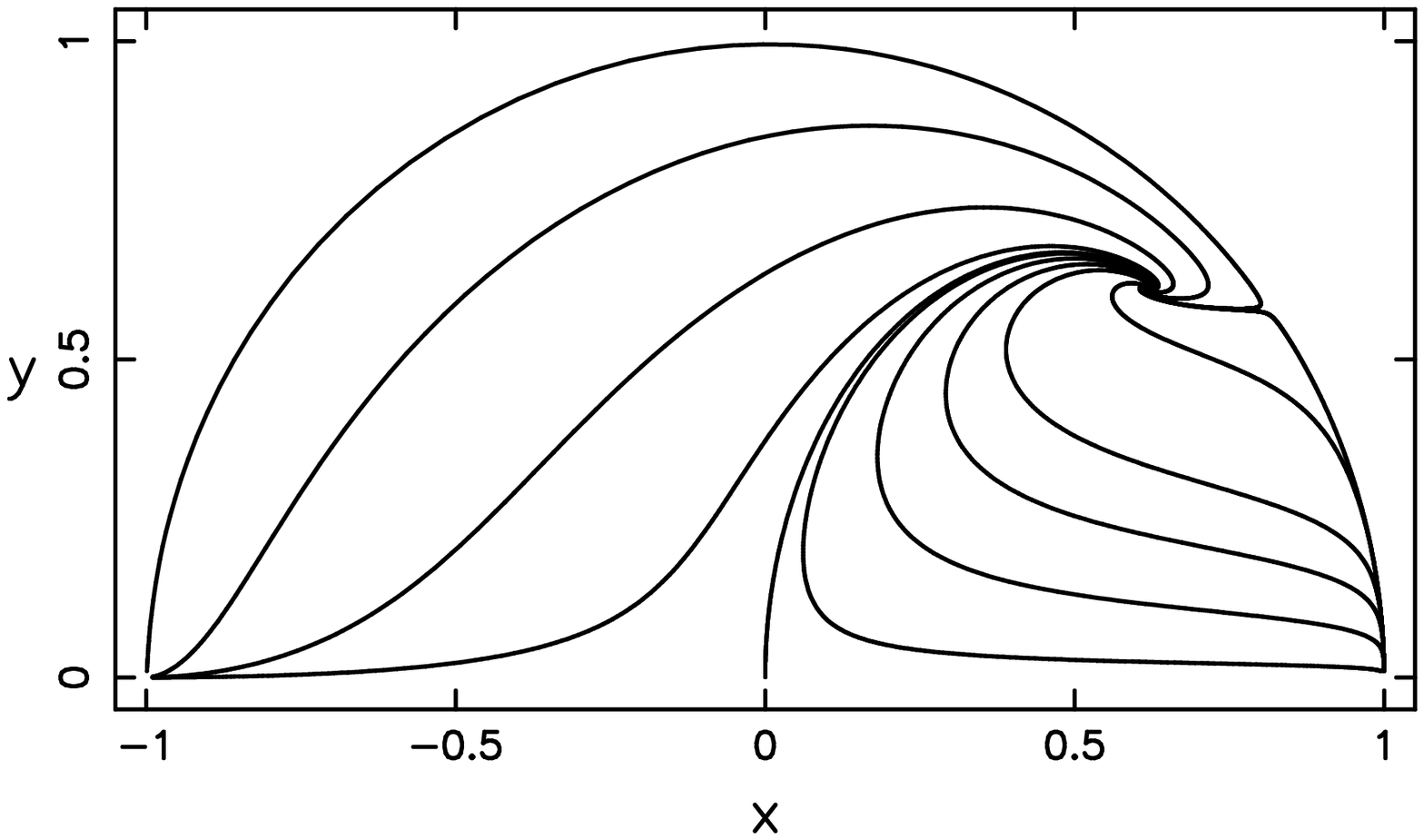}\\ 
\caption[phase2]{\label{phase2} 
The phase plane for $\gamma = 1$, $\lambda = 2$. The scalar field
dominated solution is a saddle point at $x = \sqrt{2/3}$, $y = \sqrt{1/3}$, 
and the late-time attractor
is the scaling solution with $x=y=\sqrt{3/8}$.}
\vspace*{12pt}
\leavevmode\epsfysize=5cm \epsfbox{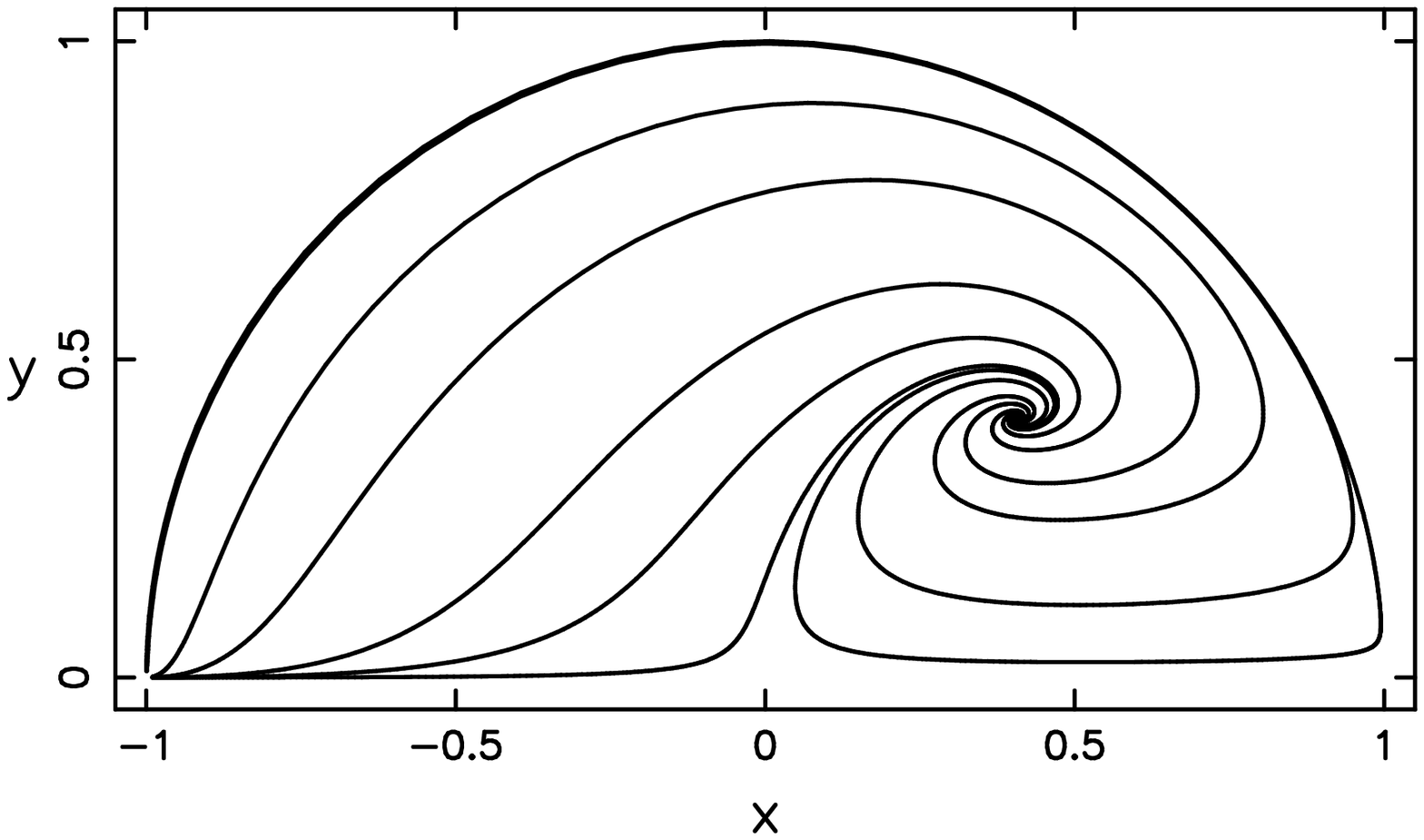}\\ 
\caption[phase3]{\label{phase3} 
The phase plane for $\gamma = 1$, $\lambda = 3$. The late-time attractor
is the scaling solution with $x=y=\sqrt{1/6}$.}
\end{figure}

The regions of $(\gamma,\lambda)$ parameter space leading to different
qualitative evolution are indicated in Fig.~\ref{regions}.
\begin{description}
\item[I.~~~]
$\lambda^2<3\gamma$. See Fig.~\ref{phase1}.\\
Both kinetic-dominated solutions are unstable nodes.
The fluid-dominated solution is a saddle point.
The scalar field dominated solution is the late-time attractor, and is 
inflationary in parameter region~Ia and non-inflationary in region~Ib.
\item[II.~~]
$3\gamma<\lambda^2<6$. See Fig.~\ref{phase2}.\\
Both kinetic-dominated solutions are unstable nodes.
The fluid-dominated solution is a saddle point.
The scalar field dominated solution is a saddle point.
The scaling solution is a stable node/spiral.
\item[III.~]
$6<\lambda^2$. See Fig.~\ref{phase3}.\\
The kinetic-dominated solution with $\lambda x<0$ is an unstable
node.
The kinetic-dominated solution with $\lambda x>0$ is a saddle point.
The fluid-dominated solution is a saddle point.
The scaling solution is a stable spiral.
\end{description}

\section{Cosmological consequences}

The possible role of scalar fields with exponential potentials with
$\lambda^2<2$ in driving an inflationary expansion of the early universe
has already received considerable attention and so we will not devote
much time to discussing this here. However we note that because the
scalar field dominated solution is the late-time attractor for
$\lambda^2<3\gamma$, the existence of such a scalar field today is ruled
out unless its energy density has been greatly suppressed relative to
the attractor value ($\Omega_\phi=1$) for most of the ``dust-dominated''
era where $\gamma=1$. Nonetheless such models have been considered as
possible `decaying cosmological constant' models~\cite{RP,FJ,LV}.

The peculiar properties of the scaling solution, which we see is the
late-time attractor for exponential potentials with $\lambda^2>3\gamma$, are
sufficiently novel to merit greater investigation. (See also
Refs.~\cite{Wetterich95,FJ}.)

The most striking possibility is that a scalar field with an exponential
potential could comprise a significant fraction of the energy density of
our universe today. Because the effective barotropic index of the
homogeneous scalar field would mimic pressureless dust, its dynamical
effect would be exactly like cold dark matter. For instance, if
$\lambda=3$ then we expect $\Omega_\phi=1/3$ today.
However the inhomogeneous field can have a different equation of state
modifying the evolution of large-scale structure in the universe, as has
recently been investigated elsewhere~\cite{FJ,LV}.

The main problem with this scenario is if the scalar field has a
significant contribution to the energy density, throughout the present
dust-dominated era, then, unlike conventional cold dark matter, it
should also have had a significant effect during the radiation-dominated
era. For $\lambda=3$ we expect $\Omega_\phi=4/9$ when
$p_\gamma=\rho_\gamma/3$.

The tightest constraint on the total energy density of the universe
comes from models of nucleosynthesis \cite{Wetterich,texas}. The primordial 
abundances of the
light elements place tight constraints on the expansion rate,
and hence the energy density, at the time of nucleosynthesis, when
$T\sim1$~MeV. If we require $\Omega_\phi < \Omega_\phi^{{\rm max}}$ at the 
time of
nucleosynthesis, then this implies 
\be
\label{nucbound}
\lambda^2 > {4 \over \Omega_\phi^{{\rm max}}} \ ,
\ee
The current upper bound on $\Omega_\phi$ at nucleosynthesis is estimated to 
be in the range $0.13$ to $0.2$ \cite{FJ}; we'll adopt the higher value to be 
conservative. Satisfying the nucleosynthesis bound requires $\lambda^2 > 20$.

Thus there is a relic abundance problem for any particle physics
theories that predict the existence of scalar fields with exponential
potentials with $\lambda^2<20$ at low energies ($T\sim1$~MeV).
Scalar fields with exponential potentials completely dominate the
energy density of the universe at nucleosynthesis for $\lambda^2<4$,
and still have an unacceptably high energy density at nucleosynthesis
for $4<\lambda^2<4/\Omega_\phi^{{\rm max}}$ unless the initial energy density 
in the
field is extraordinarily low.

To quantify exactly how small the initial energy density must be to
evade the bound Eq.~(\ref{nucbound}), we expand to first-order about the
fluid-dominated solution to find the rate at which $\Omega_\phi$ grows
away from zero --- see Appendix.  We find $\Omega_\phi\propto a^{3\gamma}$, 
which
implies that the scalar field energy density remains essentially
constant due to the large friction term in the evolution equation 
Eq.~(\ref{eq3}) as the barotropic fluid density redshifts as
$\rho_\gamma\propto a^{-3\gamma}$. Thus the scalar field acts like a
cosmological constant until $\Omega_\phi$ approaches its attractor
value. To have not reached the attractor by some time $t_{{\rm f}}$ requires 
an initial value at $t_{{\rm i}}$ satisfying
\be
\label{phii}
\Omega_{\phi} (t_{{\rm i}}) \lesssim {\rho_\gamma(t_{{\rm f}}) \over 
\rho_{\gamma} (t_{{\rm i}})} \,.
\ee

\section{The role of inflation}

The usual cosmological solution to relic abundance problems is a period of 
inflation, during which the unwanted relics have their energy density 
redshifted to a negligible value relative to the potential energy $U$ of the 
inflaton field $\sigma$.\footnote{Note that we are assuming that the field 
$\sigma$ is unrelated to the field $\phi$ which we have been discussing up 
until now.}
If inflation ends at some energy density $\rho_{\gamma}(t_{{\rm i}})\sim 
M^4$, 
then from Eq.~(\ref{phii}) we require
\be
\label{inflationbound}
\Omega_{\phi} (t_{{\rm i}}) \lesssim \left( {1\ {\rm MeV} \over M} \right)^4
\ee
for the scalar field not to have reached the scaling attractor
solution by the time of nucleosynthesis.  However we will now show
that the expected density of the scalar field at nucleosynthesis is
not significantly affected by standard models of inflation.

During conventional slow-roll inflation the inflaton field $\sigma$ has an
effective equation of state $\gamma \approx
2\epsilon/3$, where $\epsilon$ is the (non-negative) slow-roll
parameter controlling the slope of the potential~\cite{LL93}.  We can
treat $\gamma$ as effectively constant provided the approach to the
scaling attractor (determined by the largest eigenvalue for linear
perturbations) is faster than the movement of that attractor. During
slow-roll inflation this requires
\be
\label{quasib}
\left| {\epsilon' \over \epsilon} \right| \lesssim 2\epsilon \ ,
\ee
which is valid for most models, including chaotic
inflation with $U(\sigma)\propto\sigma^n$ for $n\gtrsim2$.

The inflaton-dominated solution where $\Omega_\phi=0$ is not an
attractor solution unless $\gamma=0$ and so for $\lambda^2>2\epsilon$
the attractor solution is the scaling solution with
\be
\Omega_\phi \approx {2\epsilon \over \lambda^2} \ . 
\ee
Unlike ordinary matter the energy density of a scalar field with an
exponential potential {\em does not vanish} relative to the inflaton energy 
density
during inflation, even though for $\lambda^2>2$ the exponential
potential would not have an inflationary equation of state in the
absence of the inflaton.

To evade the nucleosynthesis bound on $\lambda$ and satisfy
Eq.~(\ref{inflationbound}) requires an extremely small value of
$\epsilon$.  In slow-roll inflation $\epsilon$ must be smaller than
unity, but is not usually very small. For instance in chaotic
inflation with a potential $U(\sigma)=m^2\sigma^2/2$, 
$\epsilon \simeq 0.01$ when perturbations on the current horizon scale were 
generated, and increases to unity at the end of inflation before the
inflaton starts oscillating about the minimum of its potential. Thus
the energy density of a scalar field with an exponential potential is
not significantly reduced during chaotic inflation, and can rapidly
attain its attractor solution when the radiation-dominated era begins.

Models such as hybrid inflation~\cite{hybrid} or thermal 
inflation~\cite{therminf} are
distinctive in that $\epsilon$ can remain very small during the final
stages of inflation and thus $\Omega_\phi$ may be very small at the
attractor scaling solution. However in this case the condition
Eq.~(\ref{quasib}) for quasi-constant $\gamma$ is violated and the
attractor value for $\Omega_\phi$ approaches zero faster than the
actual solution. We effectively have $\gamma \to 0$, which corresponds
to the critical case where the scaling solution tends to the
fluid-dominated solution with $x=y=0$.

For $\gamma=0$ the largest eigenvalue for linear perturbations
vanishes (see Appendix) and we must consider higher-order perturbations about 
the
critical point to determine its stability. We find that $x=y=0$ is a
stable attractor, but that trajectories only approach this as the
logarithm of the scale factor, $N$. The late-time evolution is
given by
\be
y^2 = {\sqrt{6}\over \lambda} x \approx {1 \over \lambda^2 N} \ .
\ee
Thus even the extreme case of a cosmological constant (or constant
false-vacuum energy density) only dilutes the energy density of the
scalar field as the logarithm of the scale factor,
$\Omega_\phi\propto1/N$. Thus a model such as thermal inflation, which
is so effective at diluting the abundance of relic moduli particles 
\cite{therminf}, has a
negligible effect on the relic density of the scalar field $\phi$ as
it only lasts for a small number of $e$-foldings (typically about 15) and the
scalar field can rapidly return to its scaling solution after
inflation.
Even in the case of hybrid inflation the relic density after inflation
may be significant unless we have an extremely large number of
$e$-foldings during inflation.
To satisfy Eq.~(\ref{inflationbound}) requires 
\be
N \gtrsim {1\over\lambda^2} \left( {M \over 1\ {\rm MeV}} \right)^4
\ee
which could be of order $10^{60}$ $e$-foldings for a typical hybrid inflation 
model. Requiring so much expansion would be a significant constraint on the 
model.

\section{Conclusions}

We have presented a phase-plane analysis of the evolution of a
spatially-flat FRW universe containing a barotropic fluid plus a
scalar field with an exponential potential
$V(\phi)=V_0\exp(-\lambda\kappa\phi)$. We have shown that the energy
density of the scalar field dominates at late times for
$\lambda^2<3\gamma$. For $\lambda^2>3\gamma$ we find that the
barotropic fluid does not completely dominate and the energy density
of the scalar field remains a fixed fraction of the total density at
late times.

This leads to a relic density problem at nucleosynthesis in such
models if $\lambda^2 \lesssim 20$ . Standard models of inflation do
not significantly dilute the initial density of the exponential
potential and do not alleviate this bound. Only inflation models with
effectively constant energy density and an exponentially large number
of $e$-foldings (such as some models of hybrid inflation) would be able
to weaken this bound, unless the radiation-dominated era only begins shortly 
before nucleosynthesis, e.g.~Ref.~\cite{kination}.

We emphasize that we have assumed that there is no direct coupling
between the exponential potential and other matter. The only interaction is
gravitational. 

\section*{Acknowledgments}

E.J.C.~was supported by PPARC and A.R.L.~by the Royal
Society. We thank Pedro Ferreira for useful discussions, and acknowledge use
of the Starlink computer system at the University of Sussex.
\appendix
\section*{Stability of the critical points}

In order to study the stability of the critical points $(x_c,y_c)$ we
expand about these points 
\be 
x=x_c+u \ , \qquad y=y_c+v \ , 
\ee
which when substituted into Eqs.~(\ref{eomx}) and~(\ref{eomy}) yields, to
first-order in the perturbations, equations of motion 
\be
\left(
\begin{array}{c}
u' \\
v'
\end{array}
\right) = {\cal M} \left(
\begin{array}{c}
u \\
v
\end{array}
\right) \ .
\ee
The general solution for the evolution of linear perturbations can be
written as
\bea
u = u_+ \exp (m_+N) + u_- \exp (m_-N) \ , \\
v = v_+ \exp (m_+N) + v_- \exp (m_-N) \ ,
\eea
where $m_\pm$ are the eigenvalues of the matrix ${\cal M}$.
Thus for stability we require the real part of both eigenvalues to be
negative.

For the critical points listed in Table~1 we find:
\begin{description}
\item
{\em Fluid-dominated solution}:
\be
m_- = - {3(2-\gamma) \over 2} \ , \qquad m_+ = {3\gamma \over 2} \ .
\ee
\item
{\em Kinetic-dominated solutions}, ($x_c=\pm1$, $y_c=0$):
\be
m_- = \sqrt{{3\over2}} \left( \sqrt{6} \mp \lambda \right) \ , \qquad
m_+ = 3(2-\gamma) \ .
\ee
\item
{\em Scalar field dominated solution}: 
\be
m_- = {\lambda^2-6 \over 2} \ , \qquad
m_+ = \lambda^2 - 3\gamma \ .
\ee
\item
{\em Scaling solution}:
\be
m_\pm = - {3(2-\gamma) \over 4}
 \left[ 1 \pm \sqrt{ 1 - {8\gamma(\lambda^2-3\gamma) \over
\lambda^2(2-\gamma)} } \right] \ .
\ee
\end{description}

 
\end{document}